\documentclass[,final]{aipproc}

\layoutstyle{6x9}

\let\al=\alpha
\let\be=\beta
\let\ga=\gamma

\let\de=\delta

\let\ep=\epsilon

\let\la=\lambda

\let\si=\sigma

\let\th=\theta

\let\Ph=\Phi
\let\ph=\phi
\usepackage{amsmath}
\usepackage{amssymb}
\usepackage{amsthm}
\begin{document}

\title{SUSY Flat Directions -- to get a VEV or not?}

\classification{
                \texttt{12.60.Jv, 98.80.Cq}}

\author{Anders Basb\o ll}{address={Department of Physics \& Astronomy, University of Sussex, Brighton, BN1 9QH, United Kingdom}}

\begin{abstract}
We investigate the potential of SUSY flat directions (FDs). Large FD vacuum expectation values (VEVs) can delay thermalisation and solve the gravitino problem - if FDs decay perturbatively. This depends on how many and which directions get the VEVs. Recently the decay of the FDs have been studied with the VEVs as input. Here we look at how the VEVs come about -- statistically and analytically.  

\end{abstract}

\maketitle

\subsection{Supersymmetry, MSSM and flatness}
Supersymmetry, in which all ferminic particles get a bosonic partner and vice versa, is a very well motivated extension of the Standard Model (SM). It gives unification of all the gauge couplings at the same energy scale and it automatically removes all the high energy divergencies from Quantum Field Theory due to the equal magnitude, opposite sign contributions of fermionic and bosonic loops in the high energy limit. No superpartners have been observed. This emberrassment can be hidden by introducing R-parity, where all the known particles (including Higgs bosons) get +1, and all superpartners -1. Thus superpartners can only be created and destructed in pairs. This immediately makes the lightest superpartner (LSP) the favorite Dark Matter candidate - indeed, giving the LSP typical weak scale mass and coupling gives an energy density in the right ballpark. The scalar potential of the MSSM (the minimal supersymmetric extension of SM) consists of F-terms ($\sum_{\phi}|\partial W/\partial \phi|^2$) and D-terms ($\sum_{a}g_a^2/2|\sum_{\phi}\phi^\dagger T^a \phi|^2$) where $T^a, W$ are the gauge generators and the Superpotential respectively - and $\phi$ are the scalar fields. The renormalisable part of the superpotential is \cite{Aitchison} $W_{renorm}=y_u^{ij}U_iQ_jH_u-y_d^{ij}D_iQ_j H_d-y_e^{ij}E_iL_jH_d+\mu H_uH_d$\footnote{Superfields and their scalar part will be represented by the same symbol. $Q,L,E,U,D,H_u,H_d$ are lefthanded quarks, lefthanded leptons, righthanded charged leptons, up-type righthanded quarks, down-type righthanded quarks, positive-hypercharged Higgs, negative-hypercharged Higgs.}
where the Yukawa couplings are the same as in the Standard Model\footnote{We will choose the basis where the SUSY-breaking mass terms - not the Yukawas - are diagonal.}.

Flatness means that the potential is zero for nonzero field values - which can be seen to happen if, and only if, all D-terms and F-terms vanish individually.  This can only happen for exact (unbroken) SUSY and with no nonrenormalisable terms in the superpotential.
\subsection{Evolution of flat directions}
A catalogue of FDs where given in \cite{Gherghetta:1995dv}. FD evolution was studied in \cite{Dine:1995uk}. Giving VEVs\footnote{Flatness: hypercharge sum to zero, one up and one down weak charge balance, and a color and same anticolor balance -- lower index: generation, upper index: color, weak charge (in that order for $Q$).}
\begin{equation}
 Q_1^{1 1} =L_1^{2}=D_2^{\overline{1}}=\phi/\sqrt{3}
\end{equation}
The superfields can be multiplied and the product can be parameterised by a canonical field that experiences Hubble friction
\begin{equation}
 \chi=Q_1^{1 1}L_1^{2}D_2^{\overline{1}},\textbf{  } \chi=c\phi^m\textbf{  }(m=3),\textbf{  }\ddot{\phi}+3H\phi+V'(\phi)=0
\end{equation}
$m$ is used to keep example as general as possible. Adding nonrenormalisable terms to the superpotential ($M$ is a breaking scale: Planck/GUT/other) 
\begin{equation}
W=W_{renorm}+\sum_{n>3}\frac{\la}{M^{n-3}}\Ph^n
\end{equation}
where all possible gauge invariant and R-parity conserving terms will be allowed (and expected to be of order 1). All FDs can be lifted by such terms - either by itself $\frac{\la}{nM^{n-3}}\chi=\frac{\la}{nM^{n-3}}\phi^n$ ($n=m$ - if positive R-parity), itself squared $\frac{\la}{nM^{n-3}}\chi^2=\frac{\la}{nM^{n-3}}\phi^n$ ($n=2m$ - if negative R-parity) or a combination of fields in the flat direction with exactly one field not in the direction $\frac{\la}{M^{n-3}}\psi\phi^{n-1}$ (with respect to which the derivative then can be taken). The potential is:
 \begin{equation}
 V(\ph)=m_\ph^2|\ph|^2-cH^2|\ph|^2+\left(\frac{(Am_{3/2}+aH)\la \ph^n}{nM^{n-3}}+h.c\right)+|\la|^2\frac{|\ph|^{2n-2}}{M^{2n-6}}
\end{equation}
The masses are the ``real'' masses, whereas the terms of the same order in $\ph$ are Hubble induced terms. The F-term (last term) has order $2n-2$ because it is the derivative squared of n'th order ($W_n$ contributes to $V_{2n-2}$). The A-term (third term) is a coupling only between scalars. During inflation the induced terms dominate. If $c>0$ (50\% chance?? -- however, with minimal Kaehler potential it does NOT work), the minimum is displaced to $\left(\frac{\be HM^{n-3}}{\la}\right)^\frac{1}{n-2}$ ($\be$ order 1 constant) which could easily be of order $10^{16}GeV$, while after inflation the real masses dominate and the VEV will oscillate around a minimum of zero. The phase dependence of the potential creates Baryon number if $B-L$ is not conserved by the FD.

Sometimes flatness is described by monomials. The relation is this: $L_1L_2E_3=(\nu_e*\mu-e*\nu_\mu)*\tau^c$ is a monomial. This gives flatness to either term ($\nu_e,\mu,\tau^c$) or ($e,\nu_\mu,\tau^c$) which can be $\phi$($e^{i*\th_1},e^{i*\th_2},e^{i*\th_3}$) which keeps $D_a=0$ for all generators. Superterm $W_4\propto L_1L_2E_3N_1$ gives $|F_{N_1}|^2\propto|L_1L_2E_3|^2$ which is positive and thus lifts flatness, whereas the  A-term is $A*e^{i*\th_A}L_1L_2E_3+h.c.$ which is negative for one of the mentioned field combinations and thus chooses the minimum. 

\subsection{Cosmological consequences}

\cite{Allahverdi:2006wh} pointed out that FDs induce masses to inflaton decay products of order $g|\langle \ph\rangle|>H_I$ ($g$: gauge coupling, $H_I$: Hubble parameter during inflation). This prevents preheating\footnote{The instantanious transfer of energy from the inflaton to massless scalars through a parametric resonance.} since MSSM scalars are no longer massless. This lowers the cosmological reheating temperature (from $10^9$ to $10^3-10^7GeV$) and this avoids the gravitino problem (why we don't see any). Even, the FD could be the inflaton itself\cite{Allahverdi:2006xh}! However, \cite{Olive:2006uw} noticed that FD is only important if it lives long enough i.e. decay perturbatively - which it claimed was unlikely. Rather there would be immediate decay of the FD through quick particle production.

\subsection{Framework and particle production}

In \cite{us} we developed the framework for analysing particle production from FDs. The crucial thing is to work in the unitary gauge, where no unphysical Goldstone bosons appear. 

Writing the excitations of fields in a vector: $\Xi\equiv(\xi_1...\xi_i...\xi_n)^T$, we can write the Lagrangian as
\begin{equation}\label{LagrangianExpanded}
\mathcal{L}\supset\frac{1}{2}|\partial_{\mu}\Xi|^2-\frac{1}{2}\Xi^T
\mathcal{M}^2\Xi-\dot{\Xi}^TU\Xi+...
\end{equation} Make an orthogonal transformation
$\Xi^\prime=A\Xi$ ($A$ is orthogonal) with
\begin{equation}
\dot{A}^TA=U
\end{equation}
and the mixed kinetic term (U-term) disappears:
\begin{equation}
\mathcal{L}\supset \frac{1}{2}|\partial_{\mu}\Xi^{\prime}|^2-\frac{1}{2}\Xi^{\prime T} \mathcal{M}^{\prime 2}\Xi^{\prime}
\end{equation}
where $\mathcal{M}^{\prime 2}=A\mathcal{M}^2A^T=AB\mathcal{M}_d^2B^TA^T=C\mathcal{M}_d^{2}C^T$,
and $C=AB$. 

Non-perturbative particle production is investigated like this\cite{NilPelSor}:

Change to conformal fields $\chi_i=a\Xi^{\prime}_i$,
 where $a$ denotes the scale factor with equation of motion
\begin{equation}\label{eommany}
\ddot{\chi_i}+\Omega^2_{ij}(t)\chi_j=0
\end{equation}
where dots represent derivatives with respect to conformal time $t$,
and
\begin{equation}
\label{Omega} \Omega^2_{ij}=a^2
\mathcal{M^{\prime}}_{ij}^2+k^2\delta_{ij}
\end{equation}
where $k$ labels the comoving momentum. Using an orthogonal
time-dependent matrix $C(t)$, we can diagonalise $\Omega_{ij}$ via
$C^T(t)\Omega^2(t)C(t)=\omega^2(t)$,where $\omega$ is diagonal.

As the vacuum changes, a new set of creation/annihilation operators are required. We use 
Bogolyubov transformation with Bogolyubov coefficients $\alpha$ and
$\beta$ (matrices when more than one field).

Initially $\alpha=\mathbb{I}$ and $\beta=0$ while the system evolves as (matrix multiplication implied):
\begin{eqnarray}\label{alphadot}
\dot{\alpha} &=& -i \omega\alpha + \frac{\dot{\omega}}{2\omega} \beta - I \alpha - J\beta \nonumber \\
\dot{\beta} &=& \frac{\dot{\omega}}{2\omega} \alpha + i\omega\beta -
J \alpha - I\beta, \label{alandbe} \end{eqnarray} with the matrices \textit{I} and \textit{J} given by
\begin{eqnarray}
I&=&\frac{1}{2}\left(\sqrt{\omega}\,C^T\dot{C}\frac{1}{\sqrt{\omega}}+\frac{1}{\sqrt{\omega}}\,C^T\dot{C}\sqrt{\omega}\right)\nonumber \\
\label{jmatrix}
J&=&\frac{1}{2}\left(\sqrt{\omega}\,C^T\dot{C}\frac{1}{\sqrt{\omega}}-\frac{1}{\sqrt{\omega}}\,C^T\dot{C}\sqrt{\omega}\right).
\end{eqnarray}
Occupation number of the $i$th bosonic eigenstate reads (no summation)
\begin{equation}\label{n}
n_i(t)=(\beta^*\beta^T)_{ii}.
\end{equation}
This showes (still \cite{NilPelSor}) that not just rapidly changing eigenvalues of the Hamiltonian, but also rapidly changing eigenstates can create particles from the vacuum.

Since initially $\alpha=\mathbb{I}$ and $\beta=0$, eq.\ref{alphadot} shows that a non-vanishing matrix $J$ is a necessary condition to obtain $\dot{\beta}\neq 0$ and hence 
$n_i(t)\neq0$. In our framework we have\footnote{The last equation only holds if \textit{B} is constant in time.}
\begin{equation}\label{Gammaeq}
C^T\dot{C}=B^TA^T\dot{A}B=-B^TUB
\end{equation}
which shows that we just need to find $U$ and $\mathcal{M}$ and diagonalise the latter - with no need to make explicit transformation to the primed system.
\subsection{A single FD}
One flat direction often mentioned in the literature is $LLE$.

The potential is
 \begin{equation}
V=\frac{1}{2}\left(D_{H}^2+\sum_a D_a^2\right)\textrm{ with } 
D_{H}=\frac{g_1}{2}\sum_i q_i |\phi_i|^2\textrm{ and }
D_a =\frac{g_2}{2}\phi^\dag P^a \phi \end{equation} where $q_i$ is the
hypercharge, and $g_1,g_2$ are the hypercharge- and SU2 gauge
couplings.

We give these VEV's:
\begin{eqnarray}
&\langle\nu_e\rangle=\varphi e^{i\si _1}&\langle e\rangle=0\nonumber\\
&\langle\mu\rangle =\varphi e^{i\si _2}&\langle\nu_\mu\rangle=0\\
&\langle\tau^c\rangle=\varphi e^{i\si _3}&\nonumber
\end{eqnarray}
The Lagrangian reads
\begin{equation}\label{LagrangianLLE^c}
\mathcal{L}=\sum_{i=1}^3\frac{1}{2}|D_{\mu}\Phi_i|^2-V-\frac{1}{4}F_{\mu\nu}^2-\sum_i
\frac{1}{4}W_{\mu\nu}^{i2}
\end{equation}
with $F,W$ are hypercharge- and weak field strenth tensors and where for field $\phi_i$:
$D_i^{\mu}=\left[(\partial^{\mu}-iq_iA_0^{\mu})\de_{ij}-\sum_{a=1}^3
iP_{ij}^a A_a^\mu\right]\phi_j$ is the covariant derivative.
 $P^a$ is the $a^{th}$ Pauli-matrix. 

We get mixed kinetic terms\footnote{The gauge fields have a suppressed Lorentz index: zero - only this component matters, if the phases are assumed to change rapidly in time, not in space.}
\begin{equation}\label{LkinExpansion}
\mathcal{L}\supset
-\varphi^2A_0(\dot{\si_1}+\dot{\si_2}-2\dot{\si_3})-\varphi^2A_3(\dot{\si_1}-\dot{\si_2})
\end{equation}
- these (diagonal) gauge-VEV derivative mixtures are unphysical Nambu-Goldstone bosons (Goldstones). 
 Making a $U(1)$ gauge transformation on the multiplets ($\Phi_i$)
\begin{equation}
\Phi_i\rightarrow\Phi^{\prime}_i=e^{iq_i\la}\Phi_i
\textrm{  with  } \la=\frac{2\si_3-\si_1-\si_2}{3} \end{equation} and by making
a $SU(2)$ gauge transformation
\begin{equation}
\Phi_i\rightarrow\Phi^{\prime}_i=e^{iP^3\ga}\Phi_i
\textrm{  with } \ga=\frac{\si_2-\si_1}{2} \end{equation} we get rid of the Goldstones and the VEVs are \begin{eqnarray}\label{mod1VEV}
 \langle\nu_e\rangle&=&\varphi e^{i\si} \nonumber\\
 \langle\mu\rangle&=&\varphi e^{i\si}\\
 \langle\tau^c\rangle&=&\varphi e^{i\si}\nonumber
\end{eqnarray} with  $\si=(\si_1+\si_2+\si_3)/3$.  We write
excitations
 \begin{eqnarray}
 \nonumber
 \label{fields21}
 &\nu_e=(\varphi+\xi_2)e^{i(\si+\frac{\xi_1}{\sqrt{3}\varphi})}& e=(\xi_5+i\xi_6)e^{i\si}\nonumber\\
 &\mu=(\varphi+\xi_3)e^{i(\si+\frac{\xi_1}{\sqrt{3}\varphi})}&\nu_\mu=(\xi_7+i\xi_8)e^{i\si} \\
 &\tau^c=(\varphi+\xi_4)e^{i(\si+\frac{\xi_1}{\sqrt{3}\varphi})}.&\nonumber
 \end{eqnarray}
Among the kinetic terms we find
\begin{equation}
\mathcal{L}\supset
-\varphi\left(A_1(\dot{\xi}_6+\dot{\xi}_8)+A_2(\dot{\xi}_7-\dot{\xi}_5)\right).
\end{equation}
- again we have Goldstones - this time from the off-diagonal gauge generators.
They are removed by redefinitions
 \begin{eqnarray}\label{LLE^cDR}
 \nonumber
& \nu_e=(\varphi+\xi_2)e^{i(\si+\frac{\xi_1}{\sqrt{3}\varphi})}
& e=\frac{(\xi_5+i\xi_6)}{\sqrt{2}}e^{i\si} \nonumber\\
&\mu=(\varphi+\xi_3)e^{i(\si+\frac{\xi_1}{\sqrt{3}\varphi})}
&\nu_\mu=\frac{(\xi_5-i\xi_6)}{\sqrt{2}}e^{i\si} \\ 
 &\tau^c=(\varphi+\xi_4)e^{i(\si+\frac{\xi_1}{\sqrt{3}\varphi})}.&\nonumber
 \end{eqnarray} and we are in the unitary gauge!
We calculate $U,\mathcal{M}$, diagonalise the latter and find $J=0$ -- no particle production.

We found in \cite{us} in a toy model that particle production is propertional to the derivative of phase differences between the participating VEV fields. Here we gauged both differences away and found no particle production.

 UDD ($<u^{\overline{1}}>=\ph e^{i \si_1},<s^{\overline{1}}>=\ph e^{i \si_2},
<b^{\overline{1}}>=\ph e^{i \si_3}$) works exactly as LLE. Both phase differences are gauged away -- no particle production. In $(QQQ)_\textbf{4}L_1L_2L_3E$\footnote{$\textbf{4}$: transform as a 4 under $SU(2)$ i.e. uncontracted $SU(2)$ indices.}. (Squarks with identical $SU(2)$-charge chosen for simplicity) the VEV-fields are (can be chosen to be)
\begin{eqnarray}
 \nonumber
 \label{ABFfieldsQLQLQLE^c}
&u^{c1}=(\varphi+\xi_4)e^{i(\si_1+\frac{\xi_1}{\sqrt{7}\varphi})} 
&c^{c2}=(\varphi+\xi_5)e^{i(\si_1+\frac{\xi_1}{\sqrt{7}\varphi})}\textrm{     }
t^{c3}=(\varphi+\xi_6)e^{i(\si_1+\frac{\xi_1}{\sqrt{7}\varphi})} \nonumber\\
&\tau=(\varphi+\xi_9)e^{i(\si_1+\frac{\xi_1}{\sqrt{7}\varphi}-2\si_3-\frac{2\xi_3}{\sqrt{6}\varphi})}
&e=(\varphi+\xi_7)e^{i(\si_1+\frac{\xi_1}{\sqrt{7}\varphi}+\si_2+\frac{\xi_2}{\sqrt{2}\varphi}+\si_3+\frac{\xi_3}{\sqrt{6}\varphi})} 
 \nonumber\\
& e^c=(\varphi+\xi_{10})e^{i(\si_1+\frac{\xi_1}{\sqrt{7}\varphi})}
&\mu=(\varphi+\xi_8)e^{i(\si_1+\frac{\xi_1}{\sqrt{7}\varphi}-\si_2-\frac{\xi_2}{\sqrt{2}\varphi}+\si_3+\frac{\xi_3}{\sqrt{6}\varphi})}
\end{eqnarray}
and the no-VEV fields (notice complicated normalisation needed)

\begin{eqnarray}
 &u^{c2}=\frac{\xi_{11}+i\xi_{12}}{\sqrt{2}}e^{i\si_1}
&c^{c1}=\frac{\xi_{11}-i\xi_{12}}{\sqrt{2}}e^{i\si_1}\nonumber\\
&u^{c3}=\frac{\xi_{13}+i\xi_{14}}{\sqrt{2}}e^{i\si_1}
&t^{c1}=\frac{\xi_{13}-i\xi_{14}}{\sqrt{2}}e^{i\si_1}\nonumber\\
&c^{c3}=\frac{\xi_{19}+i\xi_{20}}{\sqrt{2}}e^{i\si_1}
&t^{c2}=\frac{\xi_{19}-i\xi_{20}}{\sqrt{2}}e^{i\si_1}\nonumber\\
& \nu_e=\left(\frac{\xi_{29}+i\xi_{30}}{2\sqrt{5}}
 -\frac{\xi_{31}+i\xi_{32}}{\sqrt{30}}\right)e^{i(\si_1+\si_2+\si_3)}\nonumber\\
&\nu_{\mu}=\left(
 \frac{\xi_{31}+i\xi_{32}}{\sqrt{\frac{6}{5}}}\right)e^{i(\si_1-\si_2+\si_3)}
\nonumber\\
&b^{c3}=\left(\frac{\xi_{27}+i\xi_{28}}{2\sqrt{\frac{1}{3}}}
 +\frac{\xi_{29}-i\xi_{30}}{2\sqrt{5}}
 +\frac{\xi_{31}-i\xi_{32}}{\sqrt{30}}\right)e^{i\si_1}&\\
 &s^{c2}=\left( \frac{\xi_{21}+i\xi_{22}}{\sqrt{\frac{3}{2}}}
 -\frac{\xi_{27}+i\xi_{28}}{2\sqrt{3}}
 +\frac{\xi_{29}-i\xi_{30}}{2\sqrt{5}}
 +\frac{\xi_{31}-i\xi_{32}}{\sqrt{30}}\right)e^{i\si_1}&
\nonumber\\
 &d^{c1}=\left(\frac{\xi_{15}+i\xi_{16}}{\sqrt{2}}
 -\frac{\xi_{21}+i\xi_{22}}{\sqrt{6}}
 -\frac{\xi_{27}+i\xi_{28}}{2\sqrt{3}}
 +\frac{\xi_{29}-i\xi_{30}}{2\sqrt{5}}
 +\frac{\xi_{31}-i\xi_{32}}{\sqrt{30}}\right)&e^{i\si_1}
\nonumber\\
&\nu_{\tau}=\left(\frac{\xi_{15}-i\xi_{16}}{\sqrt{2}}
 +\frac{\xi_{21}-i\xi_{22}}{\sqrt{6}}
 +\frac{\xi_{27}-i\xi_{28}}{2\sqrt{3}}
 -\frac{\xi_{29}+i\xi_{30}}{2\sqrt{5}}
 -\frac{\xi_{31}+i\xi_{32}}{\sqrt{30}}\right)&e^{i(\si_1-2\si_3)}.\nonumber
\end{eqnarray}
An example of nonzero entry in $J$ is $J_{4,17}=J_{3,18}=-\frac{\sqrt{3}(-\sqrt{k}+\sqrt{k+\frac{6g_2^2\varphi
^2}{k}})}{4\sqrt{10}(k^2+6g_2^2\varphi ^2)^\frac{1}{4}}
\si_3 '$.
The structure is general in the relavant $\varphi >> k$ limit:  $n\propto \sqrt{g_i\frac{\varphi}{k}}\si_i '$ ($n$: particle density, $k$: momentum of produced particle) - a huge number proportional to a VEV phase difference.

\subsection{Several FDs}

$UDD$ and $LLE$ can coexist. Combined there are 6 VEV fields, and one can only gauge 4 phase differences away (4 diagonal generators). However, it is just one phase for each direction - and they don't interact\footnote{U is block diagonal and $J=0$.}. There is no particle production. But $LLE$ and $QLD$ can also coexist with an $L$ field in common. \begin{eqnarray}
<d^{c^{1}}>=A\varphi e^{i\si _4}\nonumber\\
<s^{c\bar{1}}>=A\varphi e^{i\si _5}\nonumber\\
<\nu_e>=\sqrt{1+A^2}\varphi e^{i\si _3}\\
<\mu>=\varphi e^{i\si _2}\nonumber\\
<\tau^c>=\varphi e^{i\si _1}\nonumber
\end{eqnarray} where $A$ is the relation between the magnitude of VEVs. This gives particle production. 
\subsection{Problems with this picture}
There are problems with this picture. Monomials (or directions) are not independent. There are only 17 mass  terms -- or 20 if righthanded neutrinos($N$s) are included -- yet there are 712 (715 including $N$s)  independent monomials \cite{menew}(counted after my talk). To illustrate, $m_{(QQQ)_\textbf{4}L_1L_2L_3E_1}^2=1/7(m_{Q_1}^2+m_{Q_2}^2+m_{Q_3}^2+m_{L_1}^2+m_{L_2}^2+m_{L_3}^2+m_{E_1}^2)$ while $m_{(QQQ)_\textbf{4}L_1L_2L_2E_1}^2=1/7(m_{Q_1}^2+m_{Q_2}^2+m_{Q_3}^2+m_{L_1}^2+2*m_{L_2}^2+m_{E_1}^2)$. These are clearly not independent. Also, if$(QQQ)_\textbf{4}L_1L_2L_3E_1$ has a VEV, so has $(QQQ)_\textbf{4}L_1L_2L_2E_1$. Also, when is $QQQLLLE$ broken? From earlier arguments one could imagine that without $N$s it would be broken by itself squared - dimension 14. However, the space of $Q,L,E$ is 27(18+6+3) dimensional. It breaks the Standard Model completely, so D-terms remove 12 complex degrees of freedom (c.d.o.f.)\footnote{One real non-flat direction and one real gauge choice for each.}. So the D-flat space is 15 dimensional. $W_4$ (4th order superpotential) includes $QQQL$ and $QULE$ -- so $F_Q,F_L,F_U,F_E$ give 36 complex constraints and thus $W_4$ lifts the flat direction. This means it is lifted by the 6th order in the potential- eventhough its A-term is of much higher order. Including $N$s will give A-terms like $QQQLLLEN$ but the direction will still be lifted by $W_4$ (including $LLEN$). So the relation between flat directions and monomials has really broken down - see \cite{menew}.

\subsection{Investigation of the potential}
The potential must be investigated for the following reasons. It is very well to state that particle production is proportional to VEV phase differences. But do these differences have dynamical equations of motion to drive them? Also, the effective mass term must be negative for any direction to get a large VEV\footnote{This can be avoided by large A-term \cite{KasKaw}.}. There are 712 monomials - but also combinations thereof are gauge invariant ($LLE$, $UDD$ but also $LLE*UDD$ can get couplings). The formally flattest direction (the one lifted by highest order in W) is a combination of $Q,U,E$ (from monomials $UUUEE$, $QQQQU$, $QUQUE$). It is only lifted by $W_9$ ($V_{16}$)\cite{Gherghetta:1995dv}. Include $N$s, and it is lifted by $W_6$ ($V_{10}$). (Just add N to the mentioned monomials \cite{menew}.) Also, while \cite{Allahverdi:2006xh} claims that VEVs are in general hierarchical or flat directions independent, \cite{Olive:2006uw} claims that there will be several large ones. So we aspire to write down the general potential to 10th order (a rough count: 2.3 million couplings). 

\subsection{Normalisation and statistical treatment}
We want to count the number of couplings correctly. The $\textbf{2}$\footnote{Doublet under $SU(2)$.} made of 3 $Q$s look like $(QQQ)_{ijk}^{\al}=Q_{i}^{a \be}Q_{j}^{b \ga}Q_{k}^{c \al}\ep_{abc}\ep_{\be \ga}$ - it is 8 (not 27) dimensional \cite{Gherghetta:1995dv}. For instance $QQQ_{112}+QQQ_{121}+QQQ_{211}=QQQ_{121}-QQQ_{211}=0$ while $\frac{QQQ_{121}+QQQ_{211}-2QQQ_{112}}{\sqrt{6}}$ is a free parameter. If one let all 3 $Q_{i,j,k}$ with two $1$'s and one $2$ be standard gaussian $N(0,1)$ they will have $N(0,1/\sqrt{3})$ projected in the relevant direction. Adding the three will get us back to $N(0,1)$ - but we do want to know that there are one, not three, parameters - and it is only the same distribution when gaussianity is assumed. \cite{Gherghetta:1995dv} has 28 types of monomials (without family indices). We have found that these can be combined to $\sim$700 gauge invariant combinations of less than or exactly 10 fields - in $\sim$400 unique field combinations. (Field combination $H_uH_dLLE$ have combinations $H_uH_d+LLE$ and $H_uL+LH_dE$. The dimension even of a monomial can be nontrivial. $(QQQ)_\textbf{2}$ combined with $QU$ has neither 8*9=72 (product of dimensions)  nor 12*3=36 (12 ways to assign at least 2 different generations to 4 $Q$s and 3 generations of $U$ - but rather 54 dimentions (stated in \cite{Gherghetta:1995dv}, written down in \cite{menew}). For normalisation we will choose the antisymmetric tensors of $SU(2),SU(3)$-contractions to have norm 1 - while other linear combinations of field vectors -- ie. combination of family indices -- will be treated as if they were basis vectors (ie. if $a,b$ are products (including $SU(2),SU(3)$-contractions) of fields, $(a-b)/\sqrt{2}$ will be used as basis vector if $a+b=0$). Finally, we've chosen $1/n!$ for each superfield appering $n$ times in a product (must be superfield, since not welldefined for fields i.e. $(H_uH_d)^2=(H^+H^-)^2+2H^+H^-H_u^0H_d^0+(H_u^0H_d^0)^2$ cannot be normalised by $1$ and $1/2$ simultaniously).

\subsection{Work to do, in progress}
There is a statistical approach: Choose random couplings. Find minimum of potential with Monte Carlo methods. Try enough combinations to get a feeling of what VEVstructure is typical. There is an analytical approach: Impose symmetries. Assume common couplings ($m_{1/2},m_0, A$ and so on...). Investigate the role of the (formally) flattest direction. I work currently on both approaches.

\subsection{Conclusion}
 SUSY Flat Directions can have crucial influence on (p)reheating and offer a very nice solution to the gravitino problem, baryogenesis and even offer a ``known'' particle as a candidate for being the inflation. Preheating is a serious threat to this. The jury is still out, and the potential must be investigated thoroughly. 


\begin{theacknowledgments}
 I would like to thank the Danish taxpayers who have supported this work through The Danish Council for Independent Research | Natural Sciences and I thank Stephan Huber, Mark Hindharsh and David Bailin for useful discussions.
\end{theacknowledgments}

\end{document}